# Progress in Cavitated Front Gridline Ag Paste for PERC Solar Cell Architecture

**Donald Intal[a,*], Sandra Huneycutt[a], Abasifreke Ebong[a], Dana Hankey[b], Marshall Tibbetts[b]**

[a]Department of Electrical and Computer Engineering, University of North Carolina at Charlotte, 9201 University Blvd, Charlotte, 28223, North Carolina, USA
[b]ACI Materials Inc, 44 Castilian Dr., Goleta, 93117, California, USA

**Abstract: This paper reports on the progress made with the cavitated silver paste on passivated emitter rear contact cells (PERC) to reduce the series resistance. Thus, electrical and microscopy characterizations are used to examine these resistive losses. Fifteen 252 cm² PERC cells were fabricated for this assessment. Under standard test conditions, these cells achieved a mean conversion efficiency of 21.87% and a maximum of 22.07%. The mean measured fill factor was 79.98%, compared with a pseudo-fill factor of 83.60%, giving a mean deficit of 3.62 percentage points. Fill factor decreased with increasing area-normalized series resistance, which ranged from 0.53 to 0.83 Ω·cm², while open-circuit voltage and short-circuit current remained unvaried. These electrical trends support resistive current extraction as the principal contributor to fill-factor variation. Electroluminescence imaging revealed localized emission nonuniformity. Cross-sectional scanning transmission electron microscopy showed variation in the glass interfacial layer, while scanning electron microscopy identified micrometer and sub micrometer silver particles. Conductive atomic force microscopy after sequential removal of the bulk metal and glass revealed spatially heterogeneous current response. These observations are consistent with nonuniform electrical access at the contact, where restricted conducting pathways may concentrate on current and increase local voltage drops. The combined characterization identifies interfacial heterogeneity and the connectivity of conducting pathways as priorities for understanding and reducing extraction losses in cavitated silver metallization.**

# I. Introduction

Front-grid metallization must balance optical shading, silver consumption, and electrical resistance in crystalline silicon solar cells. Narrow silver (Ag) fingers can reduce shading and material use, provided that the printed grid maintains sufficient conductivity and effective electrical contact with silicon [1]. Cavitation has been investigated as an alternative to conventional three-roll milling for preparing front-grid Ag pastes [2], [3]. Earlier studies demonstrated its potential for finer metallization on passivated emitter and rear contact (PERC) solar cells. However the high series resistance and reduced fill factor (FF) were noted as potential limitations requiring an investigation [2], [3]. This was first investigated with the use of LECO (Laser Enhanced Contact Optimization), [4] after sintering. The study showed that the series resistance could be reduced remarkably on application of LECO. The FF recovered most in cells with initially high resistance, while pseudo-fill factor (pFF) remained comparatively stable. Contact-resistance measurements and conductive atomic force microscopy (cAFM) provided complementary evidence consistent with contact-related transport limitations. These findings motivate closer examination of the structural and electrical heterogeneity that may contribute to residual losses before LECO [5], [6].

This work evaluates those losses through electrical characterization and electroluminescence (EL) imaging of fifteen PERC solar cells (G1), interpreted alongside complementary microscopy. Measured FF, instrument-reported pFF, and diode-sensitive parameters are examined to distinguish resistive extraction losses from variation in diode response [7], [8]. EL provides spatial context for the cell-level measurements [9], [10], while cross-sectional scanning transmission electron microscopy examines the glass-containing interfacial layer, scanning electron microscopy resolves Ag-particle morphology, and cAFM maps local current response. The objective is to determine which electrical losses dominate performance variation and assess whether the observed microstructure and conduction heterogeneity support a physically consistent explanation of the remaining FF deficit.

# II. Experimental Methods

Cavitated Ag paste was evaluated as front-grid metallization on fifteen commercial size PERC architecture. The paste was screen-printed through a nominal 40 µm screen opening, dried for 2-minutes followed by sintering at 790 °C peak temperature and a belt speed of 210 in/min. The screen opening specifies the printing geometry and does not represent a measured fired-finger width. Electrical and electroluminescence measurements were then performed.

Current–voltage measurements were performed using a Sinton Instruments FCT-650 solar-cell tester under standard test conditions: an AM1.5G reference spectrum, an irradiance of 1000 W/m², and a cell temperature of 25 °C. The evaluated parameters included conversion efficiency, open-circuit voltage, short-circuit current, and fill factor (FF). Additional instrument-reported parameters included pseudo-fill factor (pFF), area-normalized series and shunt resistances, second-diode saturation-current density, and effective ideality factors at 0.1 and 1 sun. The difference between pFF and measured FF was expressed in percentage points and considered alongside series resistance when evaluating extraction losses. The reported series resistance represents an effective cell-level quantity rather than an independently measured specific Ag–Si contact resistivity.

Electroluminescence (EL) images were acquired at a forward-injection current of approximately 5.84 A and an exposure time of 400 ms. Images numbered 1–15 corresponded directly to electrical records 1–15, allowing spatial emission features to be examined alongside each cell's electrical characteristics. The images were used to assess emission uniformity and localized low-emission features; no quantitative local-resistance map was extracted from the single-current measurements.

Ag-particle morphology was examined by scanning electron microscopy. Selected lateral particle dimensions were recorded from the calibrated images. The glass-containing contact interface was examined in cross-section by high-angle annular dark-field scanning transmission electron microscopy. Five locations were marked to assess local variation in apparent interfacial thickness.

Conductive atomic force microscopy (cAFM) was used to examine local electrical conduction after sequential chemical removal of the contact layers. The sample was first etched with nitric acid ($HNO_3$) to remove the bulk Ag and expose the glass-containing interfacial layer, after which current maps were acquired to assess its spatially resolved electrical response. Subsequent etching with hydrofluoric acid (HF) removed the glass, allowing current mapping of the exposed contact region beneath the original gridline. Current maps obtained after each etching step were examined for localized conductive activity, with their respective current scales retained.

# III. Results and Discussion

The electrical characteristics of the 15 cavitated-Ag PERC cells are summarized in Table I. The cells achieved a mean conversion efficiency of 21.87 ± 0.16%, with a maximum of 22.07%. Measured FF ranged from 79.09% to 80.76%, averaging 79.98 ± 0.53%, while open-circuit voltage and short-circuit current occupied comparatively narrow ranges of 671–674 mV and 10.22–10.27 A, respectively. At fixed incident power, efficiency is proportional to the product of open-circuit voltage, short-circuit current, and FF. The larger relative variation in FF therefore accounts for much of the efficiency spread, consistent with the strong correlation between efficiency and FF ($r \approx 0.978$).

*Table 1: Electrical Characteristics of Cavitated-Ag PERC Cells (n = 15)*

| Parameter | Mean ± SD | Min. | Max. |
|---|---|---|---|
| Eff. (%) | 21.87 ± 0.16 | 21.58 | 22.07 |
| $V_{OC}$ (mV) | 673.1 ± 1.0 | 671 | 674 |
| $I_{SC}$ (A) | 10.24 ± 0.016 | 10.22 | 10.27 |
| FF (%) | 79.98 ± 0.53 | 79.09 | 80.76 |
| pFF (%) | 83.60 ± 0.14 | 83.37 | 83.78 |
| $R_S$ (Ω-cm$^2$) | 0.649 ± 0.095 | 0.53 | 0.83 |
| $R_{SH}$ (kΩ-cm$^2$) | 60.4 ± 20.4 | 7.0 | 83.8 |
| $J_{02}$ ($10^{-9}$ A/cm$^2$) | 3.68 ± 0.71 | 2.8 | 4.9 |
| $n_{0.1sun}$ | 1.089 ± 0.018 | 1.068 | 1.120 |
| $n_{1sun}$ | 0.987 ± 0.006 | 0.974 | 0.995 |
| ΔFF = pFF – FF (%) | 3.62 ± 0.54 | 2.98 | 4.63 |

The comparison between measured FF and instrument-reported pFF helps identify the electrical origin of this variation. As shown in Fig. 1(a), pFF exceeded measured FF for every cell and occupied a narrower distribution, averaging 83.60 ± 0.14%. The mean separation between pFF and FF is 3.62 percentage points. Cells 1 and 10 illustrate this distinction: their pFF values were nearly identical at 83.72% and 83.74%, whereas their measured FF values were 79.09% and 80.76%, respectively. Their corresponding area-normalized series resistances were 0.83 and 0.53 Ω·cm². Thus, similar pseudo-fill factors were accompanied by substantially different FF losses under current extraction. The inverse relationship between FF and series resistance in Fig. 1(b) ($r \approx -0.966$) supports resistive transport as the principal contributor to the observed FF variation. Because these instrument-derived parameters can share extraction inputs, their agreement does not independently locate the resistance within the device.

Diode-sensitive parameters provide additional context for the resistive-loss interpretation. Figure 1(c) shows that higher extracted second-diode saturation-current density, spanning (2.8–4.9) × $10^{-9}$ A/cm², generally accompanied lower pFF. Within a conventional two-diode model, a larger second-diode contribution increases recombination current at a given junction voltage and reduces the squareness of the pseudo-current–voltage characteristic. These differences influence the available pFF, while resistive losses further reduce the measured FF. For example, cell 5 had a pFF of 83.40%, yet its relatively low series resistance of 0.55 Ω·cm² supported a measured FF of 80.33%. Shunt leakage was also considered: cell 14 exhibited the lowest shunt resistance, 6999 Ω·cm², corresponding to approximately 0.096 mA/cm² at 0.672 V under a linear shunt approximation. Its lower shunt resistance therefore does not, by itself, establish ohmic leakage as the dominant source of its one-sun performance loss.

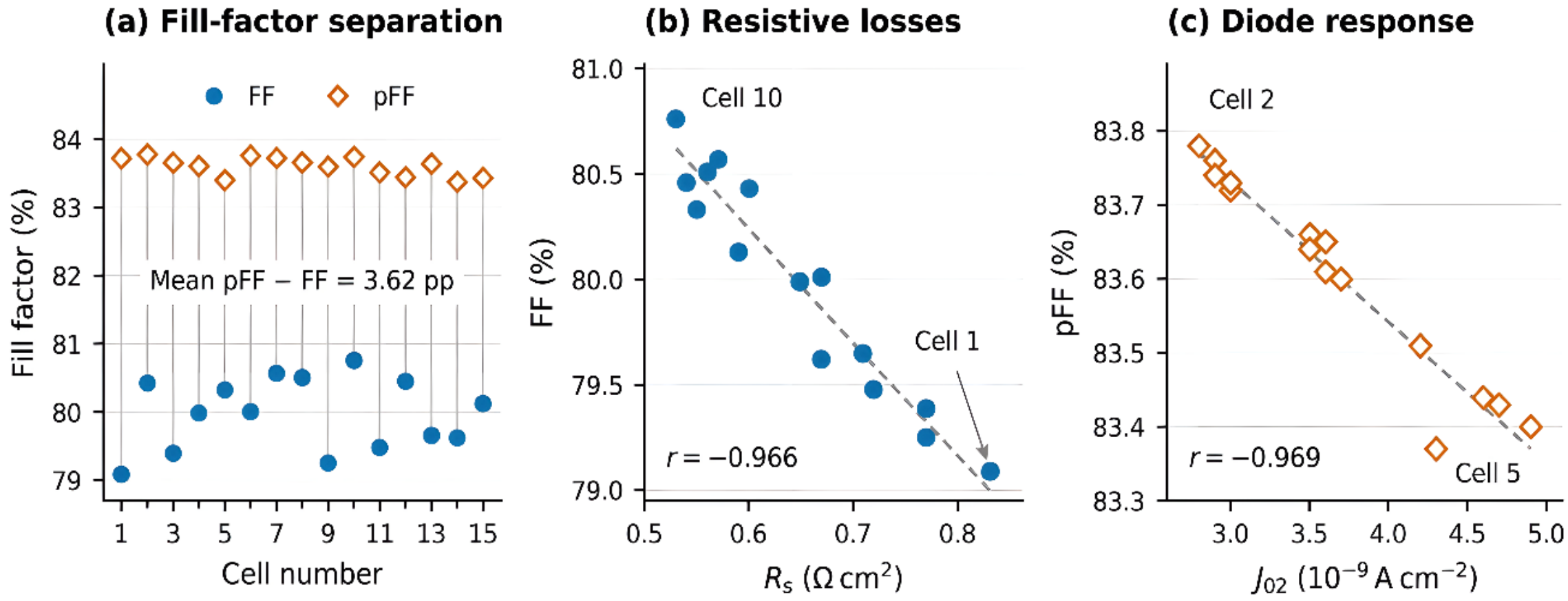

*Figure 1: (a) Measured FF and pFF by cell; (b) FF versus area-normalized series resistance; (c) pFF versus extracted $J_{02}$.*

The series-resistance contribution can be expressed through the lumped relation $V = V_j - JR_s$, where $V$ is terminal voltage, $V_j$ is junction voltage, and $J$ is the delivered current density. During power generation, the resistive voltage drop reduces the voltage available at the terminals and dissipates power per unit area proportional to $J^2R_s$, lowering the maximum power and measured FF. Current extraction involves lateral transport through the emitter, transfer across the contact interface, and conduction along the Ag grid. The continuity and spatial distribution of these pathways therefore influence the effective series resistance. Fig. 2 illustrates how restricted current transfer at the interface can concentrate current into discrete contact sites and increase local voltage drops, providing a possible physical explanation for the observed FF deficit.

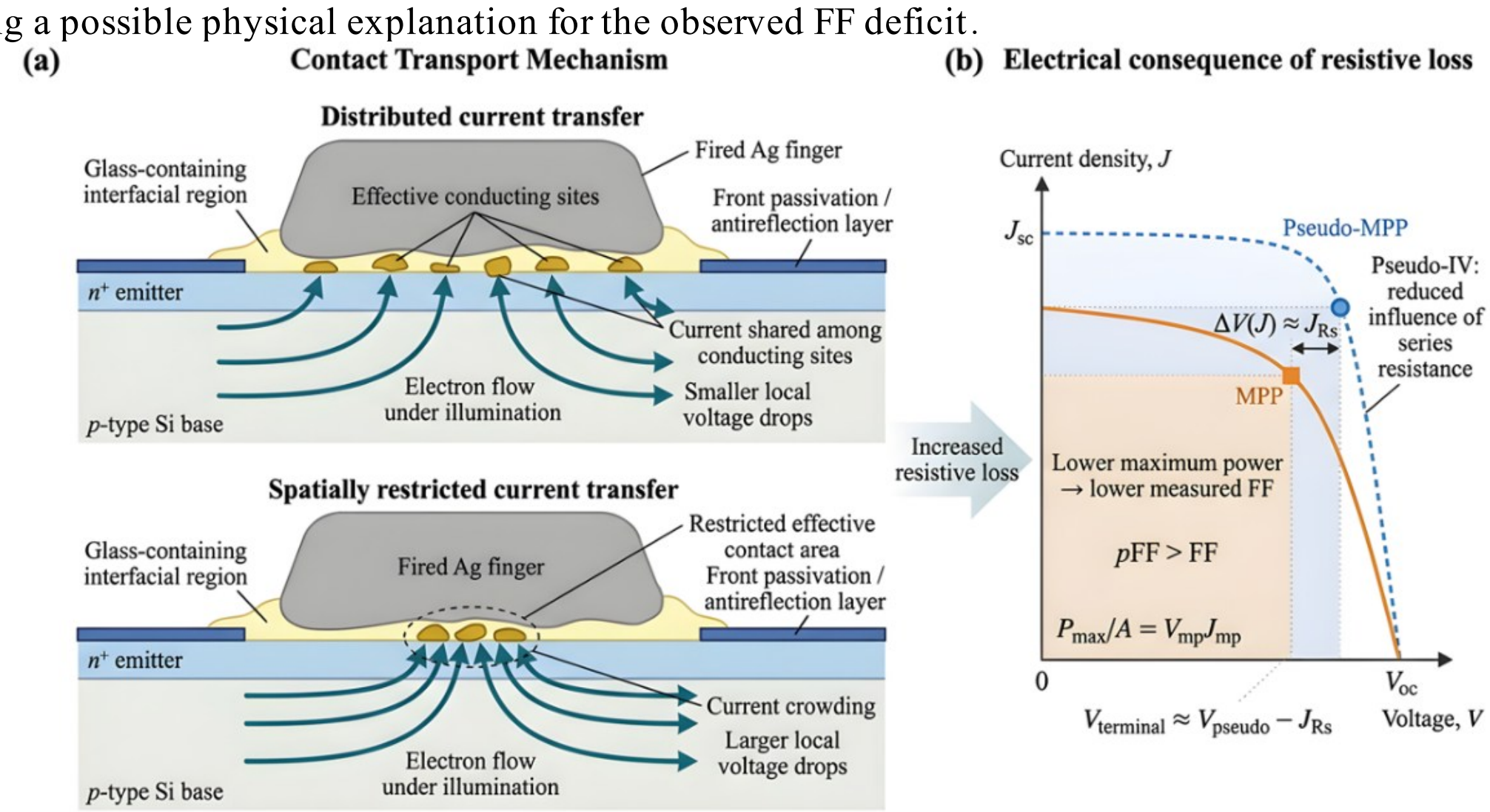

*Figure 2: Conceptual illustration of distributed versus restricted current transfer and the resulting terminal-voltage and FF losses.*

EL imaging provides spatial context for the electrical differences identified in Fig. 1. Cells 1 and 9 in Figs. 3(a) and 3(b) combined relatively high series resistance with low FF, while cell 10 in Fig. 3(c) exhibited comparatively uniform interior emission and the highest measured FF. Under forward injection and local quasi-equilibrium, EL intensity depends approximately exponentially on the local junction voltage, with additional contributions from optical and detection factors [11]. Local resistive voltage drops can therefore produce pronounced emission contrast by reducing the voltage available across the junction. Enhanced nonradiative recombination, interrupted current paths, and optical obstruction can also suppress emission, so dark features do not uniquely identify contact resistance. Cell 5 in Fig. 3(d) illustrates this distinction: its narrow, branching dark feature is consistent with a possible microcrack, but emission remains visible on both sides. Together with its relatively favorable FF and series resistance, this observation suggests that substantial electrical connectivity is retained.

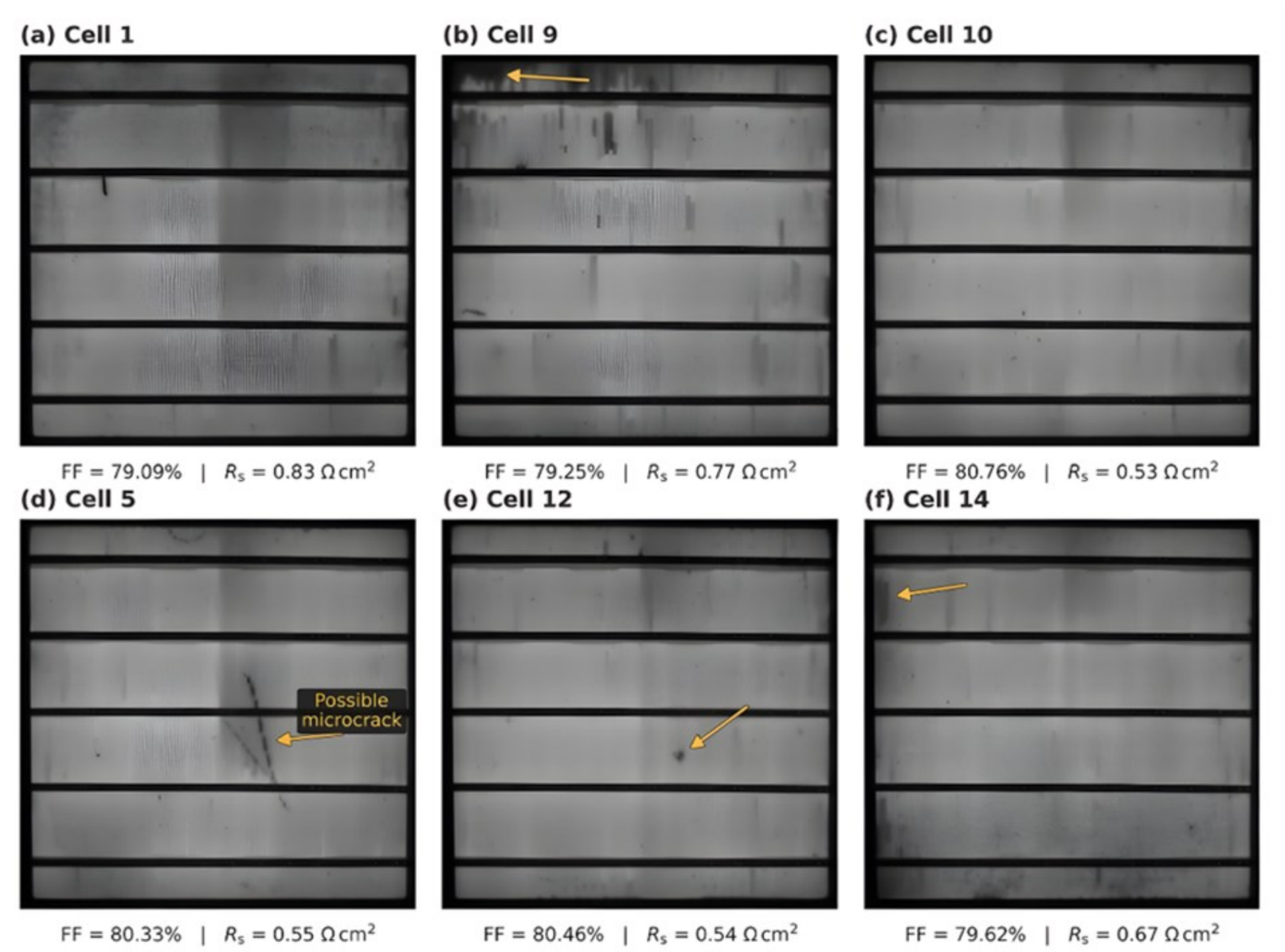

*Figure 3: Electroluminescence images of selected PERC solar cells with cavitated Ag front metallization: (a) cell 1, (b) cell 9, (c) cell 10, (d) cell 5, (e) cell 12, and (f) cell 14. Images were acquired at approximately 5.84 A with a 400 ms exposure.*

These device-level observations motivate examination of the contact microstructure, beginning with the geometry and Ag distribution at the glass-containing interface. The paired HAADF-STEM image and Ag elemental map in Fig. 4 reveal local variations in both interfacial separation and composition. The apparent separations measured along lines 1–5 were approximately 110, 118, 143, 206, and 327 nm, respectively, spanning nearly a threefold range within the displayed cross-section. A strong Ag signal originates from the bulk electrode, while localized Ag signals coincide with bright particulate features along the lower interfacial boundary. These features supports Ag-containing particles consistent with interfacial Ag crystallites reported in fired Ag–Si contacts [12]. Their location makes them possible sites for carrier transfer near the silicon surface, although their electrical contribution depends on coupling to both the semiconductor and the bulk Ag network. The elemental map establishes their composition and spatial distribution but does not establish a continuous conducting connection through the glass-containing layer.

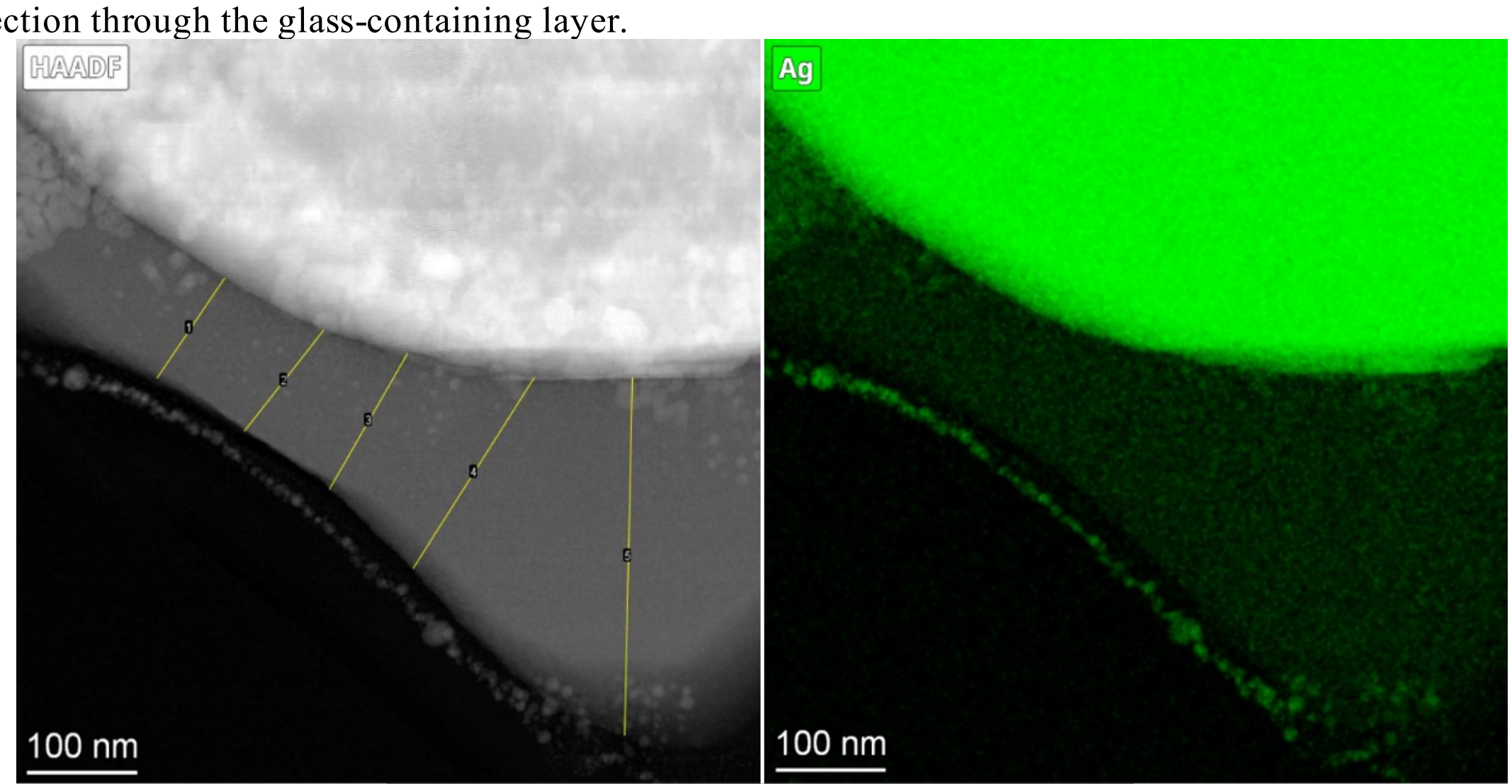

| | Line 1 | Line 2 | Line 3 | Line 4 | Line 5 |
|---|---|---|---|---|---|
| Thickness (nm) | 110.198 | 117.609 | 143.098 | 206.464 | 326.929 |

*Figure 4: Cross-sectional characterization of the cavitated-Ag/Si contact: (a) HAADF-STEM image showing five selected measurements of apparent interfacial separation and (b) corresponding Ag elemental map.*

The transport significance of this microstructure depends on the barriers encountered along electrically active paths. The apparent interfacial separations in Fig. 4 describe local geometry and should not be interpreted as uniform tunneling distances. Where sufficiently thin insulating gaps occur, tunneling probability decreases exponentially with barrier thickness, making current transfer sensitive to local variations in the intervening material. Thin glass regions, direct contact sites, and metallic inclusions separated by short insulating gaps are therefore possible preferential pathways in fired Ag contacts [12]. If only a limited fraction of the interface supports efficient transfer, carriers must travel laterally toward these sites, coupling interfacial resistance to transport through the emitter. The resulting current crowding increases local voltage gradients and provides a microscopic basis for the resistive-loss mechanism illustrated in Fig. 2. The observed geometry supports considering this mechanism without identifying which transfer pathway dominates.

In addition to carrier transfer across the contact interface, charge extraction requires conduction through the Ag grid. The SEM images in Fig. 5 examine particle morphology relevant to the formation and continuity of this conducting network. Larger Ag particles have selected lateral dimensions of 1.10–1.44 µm, while smaller particles measure 186–379 nm. The larger particles exhibit faceted or textured surfaces, and smaller particles occupy spaces between and around them. These observations describe the local particle arrangement and the coexistence of different particle sizes within the imaged regions; the annotated dimensions represent selected particles rather than a statistical size distribution.

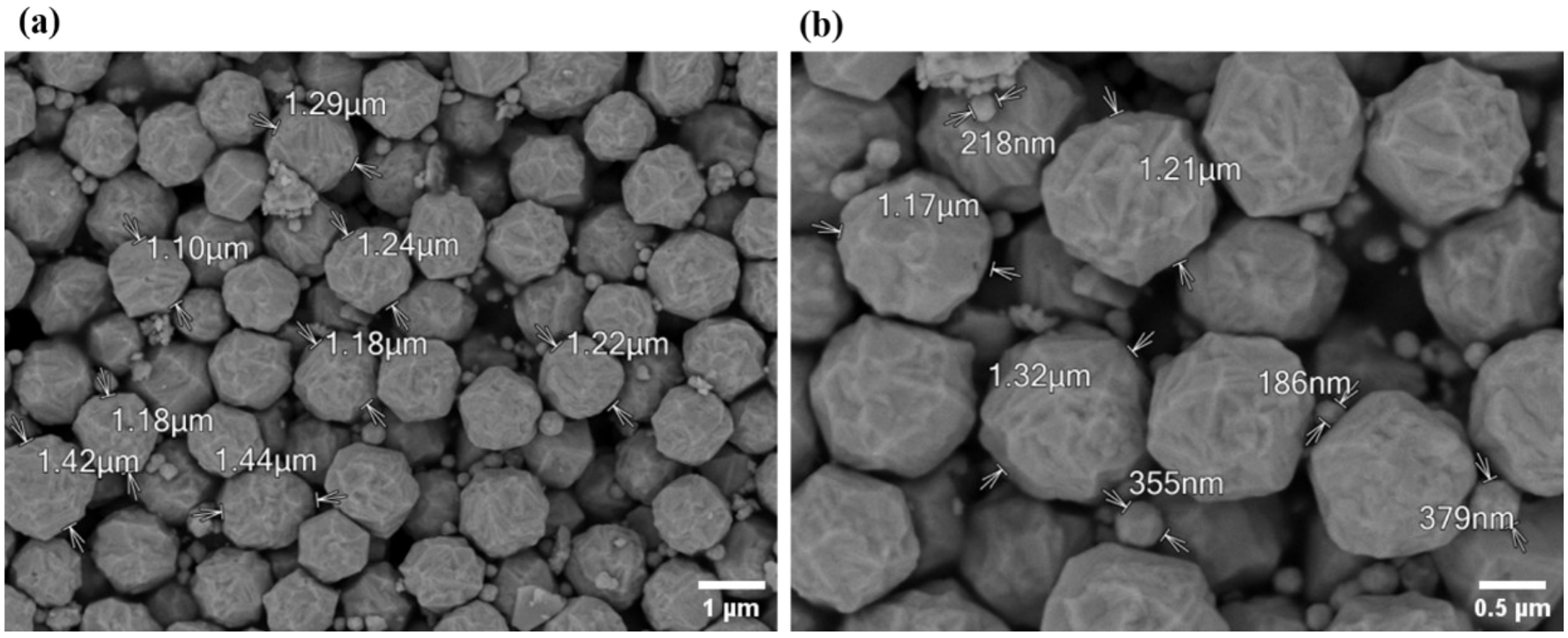

*Figure 5: Scanning electron microscopy (SEM) images of Ag-particle morphology at (a) lower and (b) higher magnification.*

The electrical significance of this arrangement depends on the connections between neighboring particles. Smaller particles may fill interstitial spaces and provide additional contact points, but their contribution to conduction depends on the development of continuous metallic pathways during firing. Sintering can enlarge interparticle necks and increase the effective conducting cross-section, whereas residual pores and narrow connections restrict current flow. For a given current, these constrictions increase local current density and resistive voltage drops. Particle morphology can therefore influence grid resistance through its effects on packing and sintering, consistent with previous investigations of solar-cell Ag powders [13]. This contribution concerns conduction along the metallic network and complements the interfacial current-transfer mechanisms considered in Fig. 4.

Complementing the structural observations in Figs. 4 and 5, the cAFM maps in Fig. 6 provide a local electrical perspective on the chemically exposed contact region. Following $HNO_3$ removal of the bulk Ag, the remaining glass-containing interface exhibits sparse bright locations within a predominantly low-current background, with a displayed current range of 0–241 pA. The localization of these responses indicates that electrical access varies across the exposed surface, consistent with the preferential transport pathways considered in Fig. 4. The measured current depends on the complete conduction path through the probe–surface contact, exposed interfacial material, and underlying semiconductor. Bright locations therefore identify comparatively greater electrical access under the measurement conditions, while the surrounding low-current regions indicate stronger restrictions somewhere along that path.

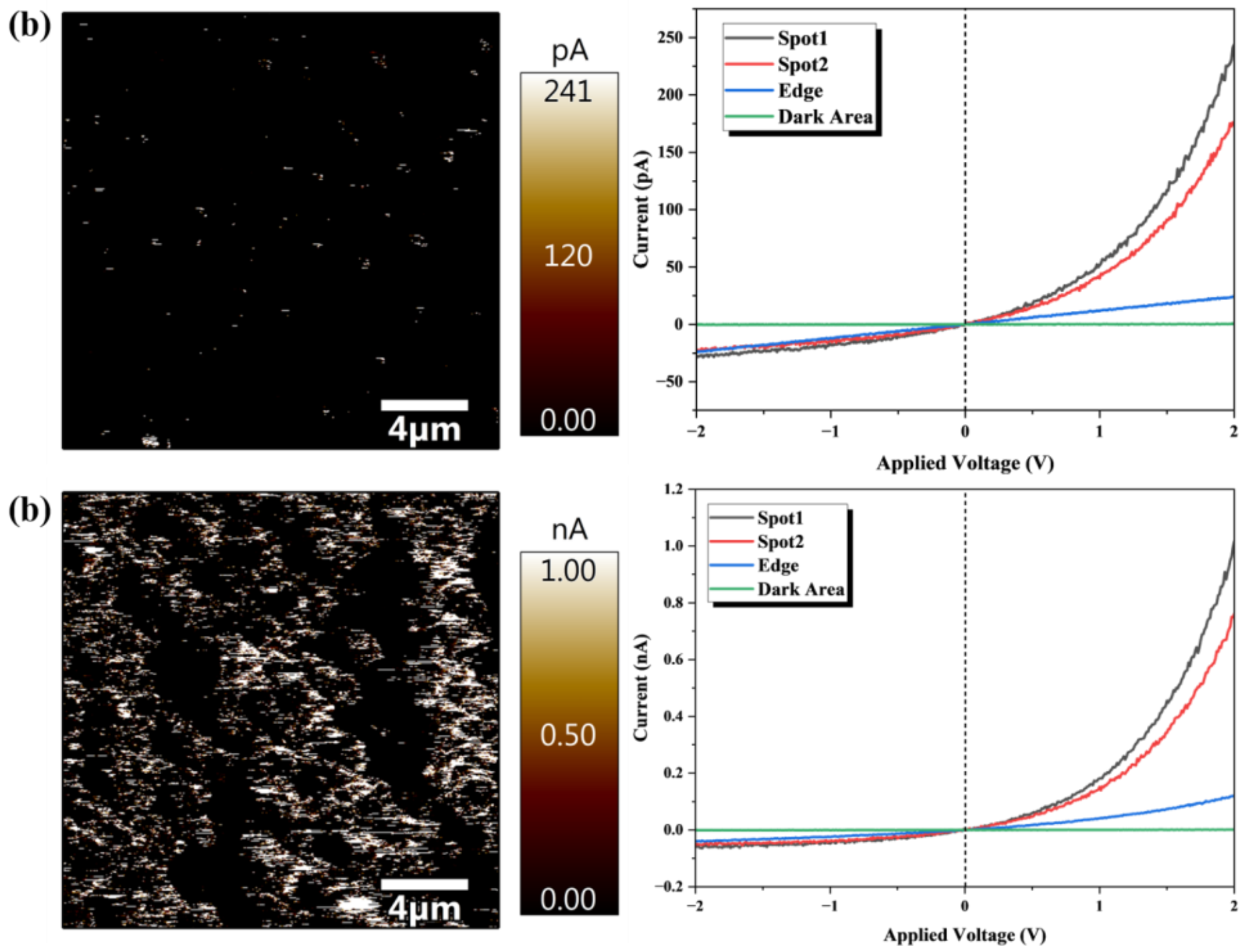

*Figure 6: Conductive atomic force microscopy (cAFM) current maps (left) and local current–voltage responses (right). The upper and lower maps show spatially heterogeneous conduction over displayed current ranges of 0–241 pA and 0–1.00 nA, respectively.*

Following HF removal of the glass, the lower map displays more extensive bright features over a current range of 0–1.00 nA. Under comparable bias and probe-contact conditions, this response is consistent with greater electrical access to the contact region beneath the original gridline after removal of the intervening glass. The different display ranges do not establish a proportional increase in conductivity, because the measured response also depends on the exposed material and probe-contact configuration. The earlier mitigation study [4] similarly reported distinct responses at conductive spots, edges, and dark regions, together with changes in current magnitude and polarity dependence following additional HF treatment. Those observations provide context for interpreting the sensitivity of local conduction to surface exposure and interfacial barriers. In the present study, the sequentially revealed surfaces add electrical evidence of heterogeneous access to the structural observations in Fig. 4. These responses characterize the prepared surfaces and do not directly quantify transport through the intact buried contact.

The electrical results support resistive current extraction as the principal contributor to FF variation. The interfacial geometry and Ag distribution in Fig. 4, particle morphology in Fig. 5, and localized current response in Fig. 6 provide complementary information about the current-transfer and grid-conduction paths illustrated in Fig. 2. Together, these observations support examining spatial restrictions in interfacial current transfer and the continuity of the Ag network as possible microscopic sources of the measured losses. Their individual contributions to the effective series resistance remain to be quantified.

# IV. Conclusion

Characterization of cavitated-Ag PERC solar cells supports resistive current extraction as the principal contributor to FF variation. The cells achieved a mean conversion efficiency of 21.87% and a maximum of 22.07%, with a mean measured FF of 79.98% compared with a pFF of 83.60%. FF decreased with increasing area-normalized series resistance over the range of 0.53–0.83 Ω·cm², while pFF, open-circuit voltage, and short-circuit current remained stable.

Cross-sectional imaging and Ag elemental mapping revealed variable interfacial geometry and localized Ag-containing particles, while SEM showed particle morphology relevant to metallic-network formation. EL showed spatial emission nonuniformity, and cAFM after sequential metal and glass removal revealed heterogeneous electrical access. These observations support considering restricted interfacial current transfer and metallic connectivity as possible microscopic sources of the remaining losses. Spatial restrictions in conducting pathways can concentrate current and increase local voltage drops, providing a physical basis for interpreting the measured FF deficit. Further work is underway to reduce series resistance, with particular attention to the suspected contact-resistance contribution. Measurements of fired-grid resistance and specific contact resistivity on matched specimens are the next characterization priority for quantifying these contributions and guiding further reductions in FF loss.